\documentclass[aps,prl,preprint,tightenlines,superscriptaddress,showpacs,byrevtex]{revtex4}

\usepackage{graphicx}
\usepackage{xspace}
\usepackage{epsfig}
\usepackage{multirow}

\newcommand{\de}{\ensuremath{\Delta E}\xspace}

\newcommand{\acp}{\ensuremath{\mathcal{A}_{CP}}\xspace}

\newcommand{\bb}{\ensuremath{B \overline{B}}\xspace}

\def\myspecial#1{}                   
\def\calL{{\mathcal L}}

\def\Mbc{M_{\rm bc}}

\begin{document}

\myspecial{!userdict begin /bop-hook{gsave 300 50 translate 5 rotate
    /Times-Roman findfont 18 scalefont setfont
    0 0 moveto 0.70 setgray
    (\mySpecialText)
    show grestore}def end}


\vspace*{-3\baselineskip}


\title{\quad\\[0.5cm] \Large
Evidence for Direct $CP$ Violation  in $B^0 \to K^+ \pi^-$ Decays}

\tighten
\affiliation{Aomori University, Aomori}
\affiliation{Budker Institute of Nuclear Physics, Novosibirsk}
\affiliation{Chiba University, Chiba}
\affiliation{Chonnam National University, Kwangju}
\affiliation{Chuo University, Tokyo}
\affiliation{University of Cincinnati, Cincinnati, Ohio 45221}
\affiliation{University of Frankfurt, Frankfurt}
\affiliation{Gyeongsang National University, Chinju}
\affiliation{University of Hawaii, Honolulu, Hawaii 96822}
\affiliation{High Energy Accelerator Research Organization (KEK), Tsukuba}
\affiliation{Hiroshima Institute of Technology, Hiroshima}
\affiliation{Institute of High Energy Physics, Chinese Academy of Sciences, Beijing}
\affiliation{Institute of High Energy Physics, Vienna}
\affiliation{Institute for Theoretical and Experimental Physics, Moscow}
\affiliation{J. Stefan Institute, Ljubljana}
\affiliation{Kanagawa University, Yokohama}
\affiliation{Korea University, Seoul}
\affiliation{Kyoto University, Kyoto}
\affiliation{Kyungpook National University, Taegu}
\affiliation{Swiss Federal Institute of Technology of Lausanne, EPFL, Lausanne}
\affiliation{University of Ljubljana, Ljubljana}
\affiliation{University of Maribor, Maribor}
\affiliation{University of Melbourne, Victoria}
\affiliation{Nagoya University, Nagoya}
\affiliation{Nara Women's University, Nara}
\affiliation{National Central University, Chung-li}
\affiliation{National Kaohsiung Normal University, Kaohsiung}
\affiliation{National United University, Miao Li}
\affiliation{Department of Physics, National Taiwan University, Taipei}
\affiliation{H. Niewodniczanski Institute of Nuclear Physics, Krakow}
\affiliation{Nihon Dental College, Niigata}
\affiliation{Niigata University, Niigata}
\affiliation{Osaka City University, Osaka}
\affiliation{Osaka University, Osaka}
\affiliation{Panjab University, Chandigarh}
\affiliation{Peking University, Beijing}
\affiliation{Princeton University, Princeton, New Jersey 08545}
\affiliation{RIKEN BNL Research Center, Upton, New York 11973}
\affiliation{Saga University, Saga}
\affiliation{University of Science and Technology of China, Hefei}
\affiliation{Seoul National University, Seoul}
\affiliation{Sungkyunkwan University, Suwon}
\affiliation{University of Sydney, Sydney NSW}
\affiliation{Tata Institute of Fundamental Research, Bombay}
\affiliation{Toho University, Funabashi}
\affiliation{Tohoku Gakuin University, Tagajo}
\affiliation{Tohoku University, Sendai}
\affiliation{Department of Physics, University of Tokyo, Tokyo}
\affiliation{Tokyo Institute of Technology, Tokyo}
\affiliation{Tokyo Metropolitan University, Tokyo}
\affiliation{Tokyo University of Agriculture and Technology, Tokyo}
\affiliation{Toyama National College of Maritime Technology, Toyama}
\affiliation{University of Tsukuba, Tsukuba}
\affiliation{Utkal University, Bhubaneswer}
\affiliation{Virginia Polytechnic Institute and State University, Blacksburg, Virginia 24061}
\affiliation{Yonsei University, Seoul}
  \author{Y.~Chao}\affiliation{Department of Physics, National Taiwan University, Taipei} 
  \author{P.~Chang}\affiliation{Department of Physics, National Taiwan University, Taipei} 
  \author{K.~Abe}\affiliation{High Energy Accelerator Research Organization (KEK), Tsukuba} 
  \author{K.~Abe}\affiliation{Tohoku Gakuin University, Tagajo} 
  \author{N.~Abe}\affiliation{Tokyo Institute of Technology, Tokyo} 
  \author{I.~Adachi}\affiliation{High Energy Accelerator Research Organization (KEK), Tsukuba} 
  \author{H.~Aihara}\affiliation{Department of Physics, University of Tokyo, Tokyo} 
  \author{K.~Akai}\affiliation{High Energy Accelerator Research Organization (KEK), Tsukuba} 
  \author{M.~Akatsu}\affiliation{Nagoya University, Nagoya} 
  \author{M.~Akemoto}\affiliation{High Energy Accelerator Research Organization (KEK), Tsukuba} 
% ...
\collaboration{The Belle Collaboration}

\begin{abstract}
We report evidence for direct $CP$ violation in the decay $B^0\to K^+\pi^-$    
 with 253~fb$^{-1}$ of data collected with the Belle detector at the KEKB 
$e^+e^-$ collider. Using 275 million $\bb$ pairs we observe a 
$B\to K^\pm\pi^\mp$ signal with $2140  \pm 53$ events.
The measured $CP$ violating asymmetry is
$\acp(K^+\pi^-) = -0.101 \pm 0.025 \mathrm{(stat)} \pm 0.005 \mathrm{(syst)}$,
corresponding to a significance of $3.9 \sigma$  including systematics.
We also search for $CP$ violation in the decays  
$B^+\to K^+\pi^0$ and $B^+\to \pi^+\pi^0$. The measured $CP$ violating 
asymmetries are $\acp(K^+\pi^0) = 0.04\pm 0.05 \mathrm{(stat)}\pm 0.02 
\mathrm{(syst)}$ and $\acp(\pi^+\pi^0)
= -0.02\pm 0.10 \mathrm{(stat)} \pm 0.01 \mathrm{(syst)}$, corresponding to   
the intervals
$-0.05 < \acp(K^+ \pi^0) < 0.13$ and $-0.18<\acp(\pi^+\pi^0)<0.14$ at  
90\% confidence level. 

\end{abstract}

\maketitle

{\renewcommand{\thefootnote}{\fnsymbol{footnote}} 
\setcounter{footnote}{0}

\normalsize

\newpage
In the Standard Model (SM), $CP$ violation arises via the interference of at
least two diagrams with comparable amplitudes but different $CP$ conserving
and violating phases. Mixing induced $CP$ violation in the $B$ sector has been
established in $b\to c\bar{c} s$ transitions \cite{2phi1,2beta}. In the SM, 
direct $CP$ violation is also expected to be sizable in the $B$ meson system 
\cite{BSS}. The first experimental evidence for direct $CP$ 
violation in $B$ mesons was shown by Belle  for the decay mode 
$B^0\to \pi^+\pi^-$ 
\cite{PIPI}. This result suggests large interference between tree and penguin 
diagrams and the existence of final state interactions \cite{FSI}. 
Recently, both
Belle \cite{belleacp} and BaBar \cite{babaracp} have reported searches for 
direct $CP$ violation  in another decay mode 
$B^0\to K^+\pi^-$, where direct $CP$ violation is also expected. 

 The $CP$ violating partial rate asymmetry is measured as: 
\begin{eqnarray}
\acp=\frac{N(\overline B \to \overline f)-N(B \to f)}
{N(\overline B \to \overline f)+N(B \to f)},
\end{eqnarray} 
where $N(\overline B \to \overline f)$ is the yield for the
$\overline{B} \to K\pi/\pi\pi$ decay and $N(B \to f)$ denotes that of the
charge-conjugate mode. 
Theoretical predictions with different
approaches suggest that $\acp(K^+\pi^-)$ could be either positive  
or negative \cite{acpth}.  
Although there are large uncertainties related to  
hadronic effects in the theoretical predictions,  results for 
 $\acp(K^+\pi^-)$ and $\acp(K^+\pi^0)$ are expected to have the same sign
and be comparable in magnitude \cite{acpth}.  
In this Letter, we report  $\acp$
measurements for $B^0\to K^+\pi^-, B^+\to K^+\pi^0$ and $B^+\to\pi^+\pi^0$ 
using 275 million $\bb$ pairs  collected
with the Belle detector at the KEKB $e^+e^-$ asymmetric-energy
(3.5 on 8~GeV) collider~\cite{KEKB} operating at the $\Upsilon(4S)$ resonance.

The Belle detector is a large-solid-angle magnetic
spectrometer that consists of a silicon vertex detector (SVD),
a 50-layer central drift chamber (CDC), an array of
aerogel threshold Cherenkov counters (ACC),
a barrel-like arrangement of time-of-flight
scintillation counters (TOF), and an electromagnetic calorimeter (ECL)
comprised of CsI(Tl) crystals located inside
a superconducting solenoid coil that provides a 1.5~T
magnetic field.  An iron flux-return located outside of
the coil is instrumented to detect $K_L^0$ mesons and to identify
muons (KLM).  The detector is described in detail elsewhere~\cite{Belle}.
Two different inner detector configurations were used. For the first sample 
of 152 million $\bb$ pairs (Set I), a 2.0 cm radius beampipe
and a 3-layer silicon vertex detector were used;
for the latter  123 million $\bb$ pairs (Set II),
a 1.5 cm radius beampipe, a 4-layer silicon detector
and a small-cell inner drift chamber were used\cite{Ushiroda}.

The $B$ candidate selection is the same as described in Ref.~\cite{btohh}.
Charged tracks are required to originate from the interaction point (IP).
Charged kaons and pions are identified using $dE/dx$
information and Cherenkov light yields in the ACC.
The $dE/dx$ and ACC information are combined to form
a $K$-$\pi$ likelihood ratio, 
$\mathcal{R}(K\pi) = \mathcal{L}_K/(\mathcal{L}_K+\mathcal{L}_\pi)$,
where $\mathcal{L}_{K}$ $(\mathcal{L}_{\pi})$ is the likelihood of
kaon (pion). Charged tracks with $\mathcal{R}(K\pi)>0.6$ are
regarded as kaons and tracks with $\mathcal{R}(K\pi)<0.4$ as pions.
Furthermore,
charged tracks that are positively identified as electrons are rejected.
The electron identification uses the information composed of $E/p$ and $dE/dx$,
shower shape, track matching $\chi^2$, and ACC light yields.
The $K/\pi$ identification efficiencies and misidentification
rates are determined from a sample of kinematically identified
$D^{*+}\to D^0\pi^+, D^0\to K^-\pi^+$ decays, where the kaon and pion from the
$D$ decay are selected in the same kinematic region as in the
$B^0\to K^+\pi^-$ decay. Table \ref{tab:kid} shows the results.
The detection efficiency for $K^-\pi^+$ is found to be $1.0\%$ greater than
that for $K^+\pi^-$; this small difference is corrected for in the $\acp$
measurement.

\begin{table}
\begin{center}
\caption{Performance of $K-\pi$ identification measured using 
$D^{*+}\to D^0\pi^+, D^0\to K^-\pi^+$ decays.}
\begin{tabular}{lcccc}
\hline\hline
   &  \multicolumn{2}{c}{Set I} & \multicolumn{2}{c}{Set II} \\ 
  & Eff. (\%) & Fake rate (\%) &  Eff. (\%) & Fake rate (\%) \\ \hline 
$K^+$ &$83.74\pm 0.18$& $5.10\pm 0.12$ & $82.41\pm 0.20$ & $6.57\pm 0.15$  \\ 
$K^-$ &$84.73\pm 0.18$ &$5.69 \pm 0.12$  & $83.26\pm 0.20$ & $7.14\pm 0.15$ \\
$\pi^+$ &$91.25\pm 0.15$ &$10.74\pm 0.15$  &$89.48\pm 0.18$ & $11.82\pm 0.17$ \\
$\pi^-$ &$90.54\pm 0.16$  &$10.09\pm 0.15$ &$88.56\pm 0.19$ &$11.57 \pm 0.17$ \\ \hline
\end{tabular}
\label{tab:kid}
\end{center}
\end{table}

Candidate $\pi^0$ mesons are reconstructed by combining two photons with
invariant mass between 115 MeV/$c^2$ and 152 MeV/$c^2$, which corresponds to
$\pm2.5$ standard deviations around the nominal $\pi^0$ mass.
Each photon is required to have a minimum
energy of 50 MeV in the barrel region ($32^\circ < \theta_\gamma < 129^\circ$)
or 100 MeV in the end-cap region ($17^\circ < \theta_\gamma < 32^\circ$ or
$129^\circ < \theta_\gamma < 150^\circ$), where $\theta_\gamma$ denotes the
polar angle of the photon with respect to the beam line.
To further reduce the combinatorial background, $\pi^0$ candidates
with small decay angles ($\cos\theta^* >0.95$) are rejected, where
$\theta^*$ is the angle between the $\pi^0$ boost direction in the
laboratory frame and its $\gamma$ daughters in the $\pi^0$ rest frame.

Two variables are used to identify $B$ candidates: the beam-constrained mass,
$M_{\rm bc} =  
\sqrt{E^{*2}_{\mbox{\scriptsize beam}} - p_B^{*2}}$, and the energy difference,
$\Delta E = E_B^* - E^*_{\mbox{\scriptsize beam}}$, where 
$E^*_{\mbox{\scriptsize beam}}$ is the beam energy and $E^*_B$ and $p^*_B$ are
the reconstructed energy and momentum of the $B$ candidates in the
center-of-mass (CM) frame. Events with 
$M_{\rm bc} > 5.20$ GeV/$c^2$ and $-0.3~{\rm GeV} < \Delta E < 0.5~{\rm GeV}$
are selected for the final analysis. 

The dominant background is from $e^+e^- \to q\bar q ~( q=u,d,s,c )$ continuum
events. To distinguish the signal from the jet-like continuum background,
event topology variables and $B$ flavor tagging information are employed.
We combine a set of modified Fox-Wolfram moments \cite{pi0pi0} into a
Fisher discriminant. The probability density function (PDF) for this
discriminant, and that for the cosine of the angle between the $B$ flight
direction and the $z$ axis, are obtained using signal and continuum
Monte Carlo (MC) events. These two variables are then combined to form
a likelihood ratio
$\mathcal{R} = {\calL}_s/({\calL}_s + {\calL}_{q \bar{q}})$,
where ${\calL}_{s (q \bar{q})}$ is
the product of signal ($q \bar{q}$) probability densities. Additional
background discrimination is provided by $B$ flavor tagging.
The  standard Belle flavor tagging algorithm \cite{tagging} gives two
outputs: a discrete variable indicating the flavor of the tagging $B$
and a MC-determined dilution factor $r$,
which ranges from zero for no flavor information to unity for unambiguous
flavor assignment. An event with a high value of $r$ ( typically
containing a high-momentum lepton) is more likely to be a $B \overline B$ event
so a looser $\mathcal{R}$ requirement can be applied. We divide the data 
into $r>0.5$ and $r<0.5$ regions.
The continuum background is reduced by applying a selection requirement on 
 $\mathcal{R}$ for events in each $r$ region of Set I and Set II 
according to the figure of merit defined as
$N_s^{exp}/\sqrt{N_s^{exp}+N_{q\bar{q}}^{exp}}$, where $N_s^{exp}$ denotes
the expected signal yields based on MC simulation and our previous branching
fraction measurements \cite{btohh} and $N_{q\bar{q}}^{exp}$ denotes the
expected $q\bar q$ yields from sideband data ($M_{\rm bc}<5.26$ GeV/$c^2$).
A typical requirement suppresses 92--99\% of the continuum background while
retaining 48--67\% of the signal.

Backgrounds from $\Upsilon(4S) \to B\overline B$ events are investigated using
a large MC sample. After the $\mathcal{R}$  requirement, 
we find a small charmless three-body background at low $\Delta E$, and 
reflections from $B^0\to \pi^+\pi^-$ decays due to $K$-$\pi$ 
misidentification.
   
The signal yields are extracted by applying unbinned two dimensional
maximum likelihood (ML) fits to the ($M_{\rm bc}$ and $\Delta E$)
distributions of the $B$ and $\overline B$ samples.
The likelihood for each mode is defined as
\begin{eqnarray}
\mathcal{L} & = & {\rm exp}\; (-\sum_{s,k,j} N_{s,k,j}) 
\times \prod_i (\sum_{s,k,j} N_{s,k,j} {\mathcal P}_{s,k,j,i}) \;\;\; 
\\
\mathcal{P}_{s,k,j,i} & = & \frac{1}{2}[1- q_i  \acp{}_j ]
P_{s,k,j}(M_{{\rm bc}\;i}, \Delta E_i),  
\end{eqnarray}
where $s$ indicates Set I or Set II, $k$ distinguishes events in the $r<0.5$ 
or $r>0.5$ regions, $i$ is the identifier of
the $i$-th event, $P(M_{\rm bc}, \Delta E)$ is the two-dimensional PDF of
$M_{\rm bc}$ and $\Delta E$, $q$ indicates the $B$ meson flavor,
$B (q=+1)$ or $\overline{B} (q=-1)$, $N_j$ is the number of events for the
category $j$, which corresponds to either signal, $q\bar{q}$ continuum,
a reflection due to $K$-$\pi$ misidentification, or
background from other charmless three-body $B$ decays.  

The yields and asymmetries for the signal and backgrounds
are allowed to float in all modes. 
Since the $K^+\pi^0$ and
$\pi^+\pi^0$ reflections are difficult to distinguish with $\Delta E$
and $M_{\rm bc}$, we fit these two modes simultaneously with a fixed
reflection-to-signal ratio based on the measured $K$-$\pi$ identification 
efficiencies
and fake rates. All the signal PDFs ($P(M_{\rm bc},\Delta E)$) are obtained
using MC simulations based on the Set I and Set II detector configurations. 
The same signal PDFs are used for events in the two different $r$ 
regions. No strong correlations between 
$M_{\rm bc}$ and $\Delta E$ are found for the  $ B\to K^+\pi^-$
signal. Therefore, its PDF is modeled by a product of a
single Gaussian for $M_{\rm bc}$ and a double Gaussian for $\Delta E$.
For the modes with neutral pions in the final state, there are correlations
between $M_{\rm bc}$ and $\Delta E$ in the tails of the signals; hence,
their PDFs are described by smoothed two-dimensional histograms.
Discrepancies between the signal peak positions in data and MC are calibrated
using $B^+ \to \overline{D}{}^0\pi^+$ decays, where the
$\overline{D}{}^0 \to K^+\pi^-\pi^0$ sub-decay is used for the modes with a
$\pi^0$ meson while $\overline{D}{}^0\to K^+\pi^-$ is used for the $K^+\pi^-$
 mode.
The MC-predicted $\Delta E$ resolutions are verified using the
invariant mass distributions of high momentum $D$ mesons. The decay mode 
$\overline{D}{}^0\to K^+\pi^-$ is used for $B^0\to K^+ \pi^-$, 
and $\overline{D}{}^0\to K^+\pi^-\pi^0$ for the modes with a $\pi^0$ in
the final state. The parameters that describe the shapes of the PDFs are
fixed in all of the fits.

The continuum background in $\de$ is described by a first or second order
polynomial while the $\Mbc$ distribution is parameterized by an
ARGUS function $f(x) = x \sqrt{1-x^2}\;{\rm exp}\;[ -\xi (1-x^2)]$, where
$x$ is $\Mbc$ divided by half of the total center of mass energy \cite{argus}.
The continuum PDF is the product of an Argus function and a polynomial, where  
$\xi$ and the coefficients of the polynomial are free parameters. These 
free parameters are $r$-dependent.
 A large MC sample is used to investigate the background 
from charmless $B$ decays and a smoothed two-dimensional histogram is taken
as the PDF.
The functional forms of the PDFs are the same for the $B$ and
$\overline{B}$ samples.  

\begin{figure}
\hspace{-1.0cm}
\epsfig{file=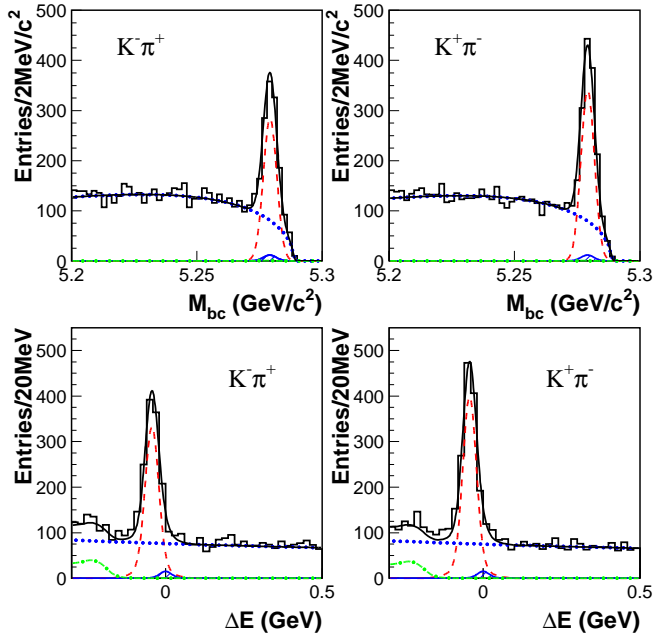,width=3.5in} 
\caption{$M_{\rm bc}$ (top) and $\Delta E$ (bottom) distributions for
${\overline B} ^0\to K^-\pi^+$ (left) and $B^0\to K^+\pi^-$ (right)
candidates. The histograms
represent the data, while the curves represent the various components from
the fit: signal (dashed), continuum (dotted), three-body $B$ decays
(dash-dotted), background from  mis-identification (hatched),
and sum of all components (solid).}
\label{fig:kpi}
\end{figure}

The efficiency of particle identification is slightly
different for positively and negatively charged particles;
consequently the raw number of asymmetry in Eq. 3 no longer gives $\acp$
correctly and must be corrected.
For the $K^+\pi^0$ and $\pi^+\pi^0$ modes, this raw asymmetry
can be expressed as:
\begin{eqnarray}
{\cal A_{CP}^{\rm raw}} = \frac{{\cal A}_\epsilon + {\cal A}_{CP}}
{1 + {\cal A}_\epsilon {\cal A}_{CP}},
\end{eqnarray} 
where $\acp$ is the true partial rate asymmetry and the efficiency asymmetry 
${\cal A}_\epsilon$ is the efficiency difference between
$K^- (\pi^+)$ and $K^+ (\pi^-)$ divided by the sum of their efficiencies. 
The situation is more complicated for 
the $K^+\pi^-$ mode because, in addition to the bias due to the efficiency 
difference between $K^-\pi^+$  and $K^+\pi^-$, a $K^-\pi^+$ signal event 
can be misidentified as a $K^+\pi^-$ candidate and dilute $\acp$.
The efficiency asymmetry results in an $\acp$ bias of +0.01 
while the small dilution factor due to double misidentification reduces  
$\acp$ by a factor of 0.99. These effects are included in the raw asymmetry
correction and their errors are included in the systematic uncertainty.
 
\begin{figure}
\hspace{-1.0cm}
\epsfig{file=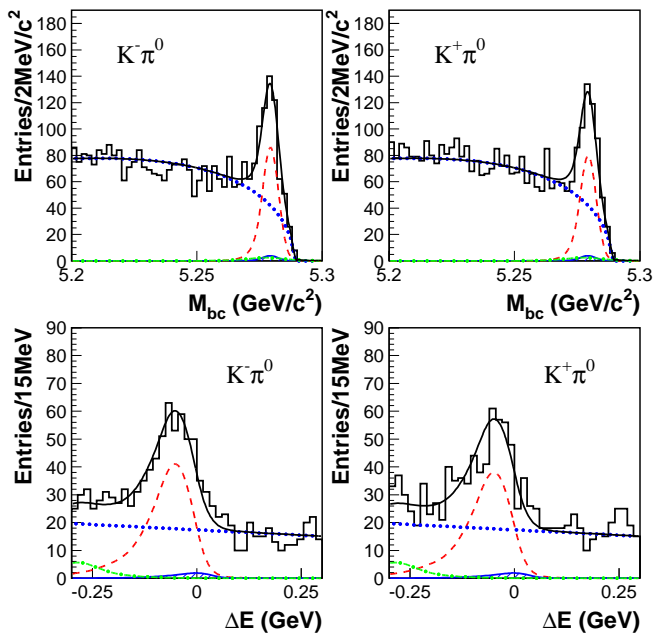,width=3.5in} 
\caption{$M_{\rm bc}$ (top) and $\Delta E$ (bottom) distributions for
$B^-\to K^-\pi^0$ (left) and $B^+\to K^+\pi^0$ (right)
candidates. The curves are described in the caption of Fig. \ref{fig:kpi}.} 
\label{fig:kpi0}
\end{figure}
   
Table \ref{tab:acp} gives the signal yields and $\acp$ values for each mode.
The asymmetries for the background components are consistent with zero within
errors.
Projections of the fits are shown in Figs.~\ref{fig:kpi}-\ref{fig:pipi0}.
The systematic errors from fitting are estimated  
from the deviations in $\acp$ after varying each parameter of the 
signal PDFs by one standard deviation. The uncertainty in modeling 
the three-body background is studied by excluding the low $\Delta E$ region
($<-0.12$ GeV) and repeating the fit. Systematic uncertainties due to  
particle identification are estimated  by checking the fit after varying 
the $K/\pi$ efficiencies and fake rates by one standard deviation. 
At each step, the deviation in $\acp$ is added in
quadrature to provide the systematic errors, which are   
less than 0.01 for all modes. A possible bias from the fitting procedure is  
checked in MC and a bias due to the $\mathcal{R}$ cut is investigated 
using the  $B^+\to \overline{D}{}^0\pi^+$ samples. No significant bias is 
observed.  
The systematic uncertainties due to the detector bias  are obtained using the 
fit results for the continuum background listed in Table \ref{tab:acp}. 
The final systematic errors are then obtained by quadratically summing the
errors due to the detector bias and the fitting systematics.

\begin{table}
\begin{center}
\caption{Fitted signal yields, $\acp$ results and background asymmetries for
 individual modes.}
\begin{tabular}{lrcc}
\hline\hline
~Mode~ & ~~~Signal Yield & $\acp$ & Bkg $\acp$\\
\hline
~$K^\mp\pi^\pm$ &$2140\pm 53$ & $-0.101\pm0.025\pm0.005$~ &$-0.001\pm 0.005$\\
~$K^\mp\pi^0$ &$728\pm 34$  &$0.04\pm 0.05\pm 0.02$& $-0.02\pm 0.01$ \\
~$\pi^\mp\pi^0$ &$315\pm 29$ &$-0.02\pm 0.10\pm 0.01$  & $-0.01\pm 0.01$ \\
\hline\hline
\end{tabular}
\label{tab:acp}
\end{center}
\end{table}

The partial rate asymmetry $\acp(K^+\pi^-)$ is found to be
$-0.101\pm 0.025\pm 0.005$, which is $3.9\sigma$ from zero.
The significance calculation includes the effects of systematic uncertainties.
Our result is consistent with the value reported by BaBar,
$\acp(K^+\pi^-) = -0.133 \pm 0.030 \pm 0.009$ \cite{babaracp}.
The combined experimental result has a  significance greater than $5 \sigma$, 
indicating that direct $CP$ violation in the $B$ meson system is established. 
Our measurement of  
$\acp(K^+\pi^0)$ is consistent with no asymmetry; the central value is
$2.4 \sigma$ away from $\acp(K^+\pi^-)$. If this result is confirmed with
higher statistics, the difference  may be due to the contribution
of the electroweak penguin diagram or other mechanisms \cite{anom}. 
No evidence of
 direct $CP$ violation is observed in the decay $B^+ \to \pi^+\pi^0$.
We set  90\% C.L. intervals $-0.05< \acp(K^+ \pi^0) < 0.13$ and 
$-0.18 < \acp(\pi^+ \pi^0)< 0.14$.

\begin{figure}
\hspace{-1.0cm}
\epsfig{file=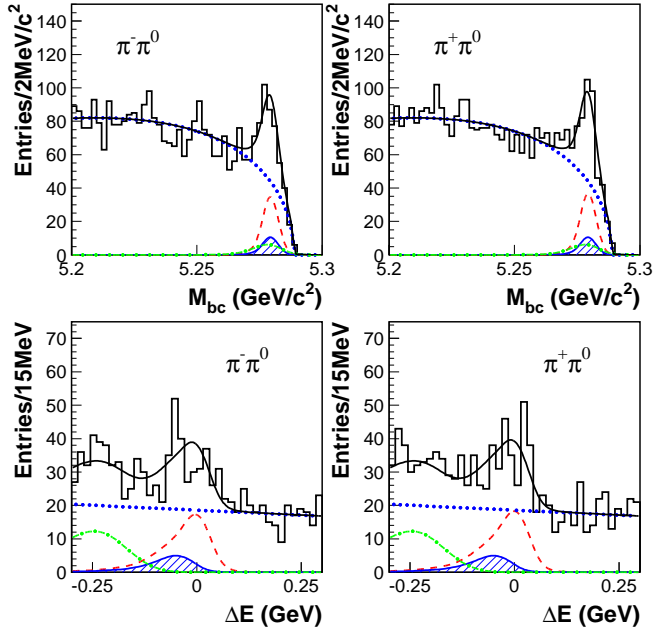,width=3.5in} 
\caption{$M_{\rm bc}$ (top) and $\Delta E$ (bottom) distributions for
$B^-\to \pi^-\pi^0$ (left) and $B^+\to \pi^+\pi^0$ (right)
candidates. The curves are described in the caption of Fig. \ref{fig:kpi}. } 
\label{fig:pipi0}
\end{figure}

We thank the KEKB group for the excellent
operation of the accelerator, the KEK Cryogenics
group for the efficient operation of the solenoid,
and the KEK computer group and the NII for valuable computing and
Super-SINET network support.  We acknowledge support from
MEXT and JSPS (Japan); ARC and DEST (Australia); NSFC (contract
No.~10175071, China); DST (India); the BK21 program of MOEHRD and the
CHEP SRC program of KOSEF (Korea); KBN (contract No.~2P03B 01324,
Poland); MIST (Russia); MESS (Slovenia); NSC and MOE (Taiwan); and DOE
(USA). 


\end{document}